\documentclass[twocolumn,aps,prl]{revtex4}
\usepackage[latin9]{inputenc}
\usepackage{float}
\usepackage{amsmath}
\usepackage{amssymb}
\usepackage{graphicx}

\makeatletter
\@ifundefined{textcolor}{}
{%
 \definecolor{BLACK}{gray}{0}
 \definecolor{WHITE}{gray}{1}
 \definecolor{RED}{rgb}{1,0,0}
 \definecolor{GREEN}{rgb}{0,1,0}
 \definecolor{BLUE}{rgb}{0,0,1}
 \definecolor{CYAN}{cmyk}{1,0,0,0}
 \definecolor{MAGENTA}{cmyk}{0,1,0,0}
 \definecolor{YELLOW}{cmyk}{0,0,1,0}
}

\usepackage{graphics}
\usepackage{setspace}
\usepackage{array}
\usepackage{textcomp}
\usepackage{rotating}
\usepackage{bm}
\usepackage{wrapfig}

\newcommand{\vex}[1]{\bm{\mathrm{#1}}}

\newcommand{\bsub}{\begin{subequations}}
\newcommand{\esub}{\end{subequations}}

\newcommand{\be}{\begin{equation}}
\newcommand{\ee}{\end{equation}}
\newcommand{\bea}{\begin{eqnarray}}
\newcommand{\eea}{\end{eqnarray}}

\makeatother

\begin{document}

\title{Irradiated three-dimensional Luttinger semimetal: A factory for engineering
Weyl semimetals}

\author{Sayed Ali Akbar Ghorashi, Pavan Hosur, Chin-Sen Ting}

\affiliation{Texas Center for Superconductivity and Department of Physics, University
of Houston, Houston, Texas 77204, USA}
\begin{abstract}
We study the interaction between elliptically polarized light and
a three-dimensional Luttinger semimetal with quadratic band touching
using Floquet theory. In the absence of light, the touching bands
can have the same or the opposite signs of the curvature; in each
case, we show that simply tuning the light parameters allows us to
create a zoo of Weyl semimetallic phases. In particular, we find that
double and single Weyl points can coexist at different energies, and
they can be tuned to be type I or type II. We also find an unusual
phase transition, in which a pair of Weyl nodes form at finite momentum
and disappear off to infinity. Considering the broad tunability of
light and abundance of materials described by the Luttinger Hamiltonian,
such as certain pyrochlore iridates, half-Heuslers and zinc-blende
semiconductors, we believe this work can lay the foundation for creating
Weyl semimetals in the lab and dynamically tuning between them. 
\end{abstract}
\maketitle
\textit{Introduction}.- Topological phases of matter have attracted
tremendous interest since the discovery of topological insulators.
Topological protection of their edge and surface states is the hallmark
of these systems, and leads to applications ranging from quantum computation
to robust transport and exotic superconductivity \cite{1,2}. In contrast
to topological insulators, which are gapped phases of matter like
most topological phases, it has been shown recently that gapless phases
of matter can be topological as well \cite{3,4,5,6,7,8,9}. Among
them, Weyl semimetals (WSMs) have been particularly attractive due
to their unconventional properties such as the chiral anomaly \cite{11,12},
negative magnetoresistance \cite{10,11} and anomalous Hall effect
\cite{5,8}. Experimental observation of these phases in TaAs \cite{11,20,21}
and photonic crystals \cite{22} has ignited further interest in exploring
these systems.

Very recently, new types of WSMs, namely, type-II and multi-WSMs were
also discovered \cite{23,mwp1,mwp2,mwp3}. The defining feature of
type-II Weyl points is that the dispersion around them is strongly
anisotropic, such that the slope changes sign along some directions.
As a result, the Weyl nodes become the touching points between electron
and hole Fermi surfaces, and result in properties different from those
of type\textendash I WSMs. For example, there are indications that
the chiral anomaly depends on the relative direction of the magnetic
field and the tilt of the cone, but the issue is still under debate
\cite{23,chiral2}. Moreover, unlike in type-I WSMs, the anomalous
Hall effect can survive in type-II WSMs under certain conditions even
when the nodes are degenerate \cite{24}. On the other hand, multi-WSMs
occur when the monopole charges of Weyl points are higher than $1$,
and can be either type-I or type-II \cite{,mwp1,mwp2,mwp3}. In general,
the search for Weyl semimetallic phases has been a vigorous field
of research lately, and proposals have been put forth to engineer
these phases in a tunable way by shining light on Dirac semimetals
\cite{29,30}, band insulators \cite{31}, stacked Graphene \cite{32},
line-nodal semimetals \cite{30,33} and crossing-line semimetals \cite{rec2,rec3}.
Finally, proposals have been made to create tunable WSMs in pyrochlore
iridates with Zeeman fields. \cite{44,rec1}.

\begin{figure}[H]
\centering \includegraphics[width=0.37\textwidth,height=0.37\textwidth]{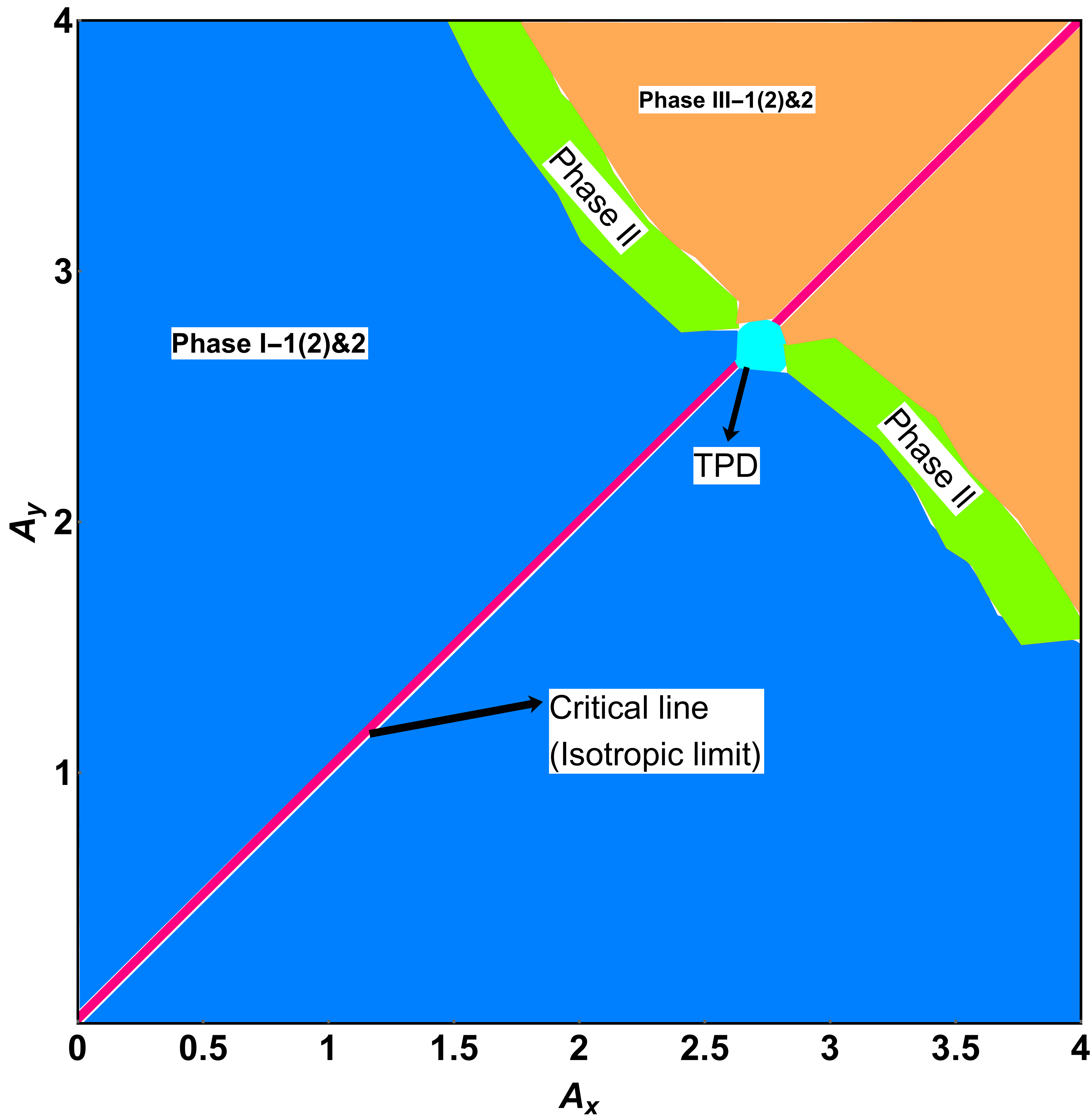}
\caption{Phase diagram for 3D Floquet Luttinger semimetal. Critical line (Isotropic
limit): diagonal red line with two lower ($w_{l}$) and two higher
($w_{h}$) Weyl points on the $k_{z}$ axis; phase I: $4w_{l}+2w_{h}$
, blue, where for bands bending oppositely (similarly), the $4w_{l}$
are type-I (type-II) denoted by $phaseI-1(2)\&2$. The notation 1(2)
denotes type-I (type-II) Weyl nodes in the respective phase. The first
number is the type of lower and the second number after $\&$ indicates
the type of the higher nodes. Phase III: $4w_{h}+2w_{l}$, orange,
where for bands bending oppositely (similarly), the $2w_{l}$ are
type-I (type-II) and denoted by $phaseIII-1(2)\&2$. Phase II, green,
shows the transient phases between phase I and III where the flat
bands in the $k_{z}$-direction ($k_{x}-k_{y}$ plane) for bands bending
in opposite (same) directions, as well as merging and splitting of
lower and upper nodes in $k_{y}-k_{z}$ and $k_{x}-k_{z}$ planes,
respectively, occur. \char`\"{}TPD\char`\"{} denotes the triply degenerate
point, which exists only for circular light.}
\end{figure}

In this work, we expand the horizons for creating tunable WSMs, by
computing the band structure of a three-dimensional Luttinger semimetal
with quadratic band touching irradiated by elliptically polarized
light using Floquet theory. We find Weyl nodes of different charges
($\pm1$ and $\pm2$), Weyl nodes of different types (type-I and type-II),
and several phases which contain more than one class of Weyl nodes.
We also stumble upon a situation where a pair of Weyl nodes form at
infinity, and rapidly come in and merge with other nodes at finite
$\mathbf{k}$. In a regularized lattice model, this pair would form
at the edge of the Brillouin zone. Crucially, given the bare band
structure, all these phases can be accessed by simply changing the
properties of the light, making this system highly tunable. Fig. 1
summarizes the results of this paper. We expect these results to hold
for real systems described by the Luttinger Hamiltonian \cite{34},
such as the zinc-blend semiconductors GaAs, HgTe, $\alpha$-Sn etc.
and a class of pyrochlore iridates \cite{35,36,37,38} studied recently.

\textit{Model and Formalism}.- We begin with an isotropic version
of the Luttinger Hamiltonian \cite{34}, 
\begin{align}
H=\frac{1}{2}\int_{\vex{k}}c^{\dagger}(\vex{k})\left((\lambda_{1}+\frac{5}{2}\lambda_{2})k^{2}-2\lambda_{2}(\vex{J}.\vex{k})^{2}-\mu\right)c(\vex{k}),
\end{align}
where $\lambda_{1,2}$ are positive constants, $\vex{k}=\{k_{x},k_{y},k_{z}\}$,
$c(\vex{k})=({c_{3/2\vex{k}},c_{1/2\vex{k}},c_{-1/2\vex{k}},c_{-3/2\vex{k}}})^{T}$,
$\vex{J}=\{J^{x},J^{y},J^{z}\}$ are effective spin-3/2 operators,
and $c_{m\mathbf{k}}$ denotes a fermion annihilation operator with
momentum\textbf{ $\mathbf{k}$ }and $J_{z}$ quantum number $m$.
The energy dispersions are $E(k)=(\lambda_{1}\mp2\lambda_{2})k^{2}-\mu$
for the $j=3/2$ and the $j=1/2$ bands, respectively. Time-reversal
and inversion symmetries ensure that the four bands come in doubly
degenerate pairs due to Kramer's theorem. The degenerate pairs of
bands curve the same (opposite) way for $\lambda_{2}<2\lambda_{1}$
($\lambda_{2}>2\lambda_{1}$), as depicted in Fig. 2. When both bands
bend the same way, Eq. (1) is widely used to model heavy- and light-hole
bands in zinc-blende semiconductors \cite{38}. Many properties of
such a dispersion have been studied in the literature, including a
recent study on the realization of fully gapped topological superconductivity
with \emph{p}-wave pairing which has states with exotic cubic and
linear dispersions coexisting on the surface \cite{39,40}. On the
other hand when bands bend oppositely, the above model is relevant
for certain pyrochlore iridates as well as for some doped half-Heusler
alloys such as LaPtBi \cite{41,42,43}. Various aspects of this scenario
have been explored as well, such as the phase diagram in the presence
of electronic interactions \cite{44}, the effect of anisotropy \cite{45}
and superconductivity \cite{46,bitanandme}. Systems with higher effective
spins and winding numbers have also attracted interest in the context
of multi-weyl phases \cite{rec4} and the investigation of the spin
quantum Hall plateau transition on the surface of topological superconductors
with general winding numbers \cite{SQHP}.\\
 
\begin{figure}[H]
\includegraphics[width=0.25\textwidth]{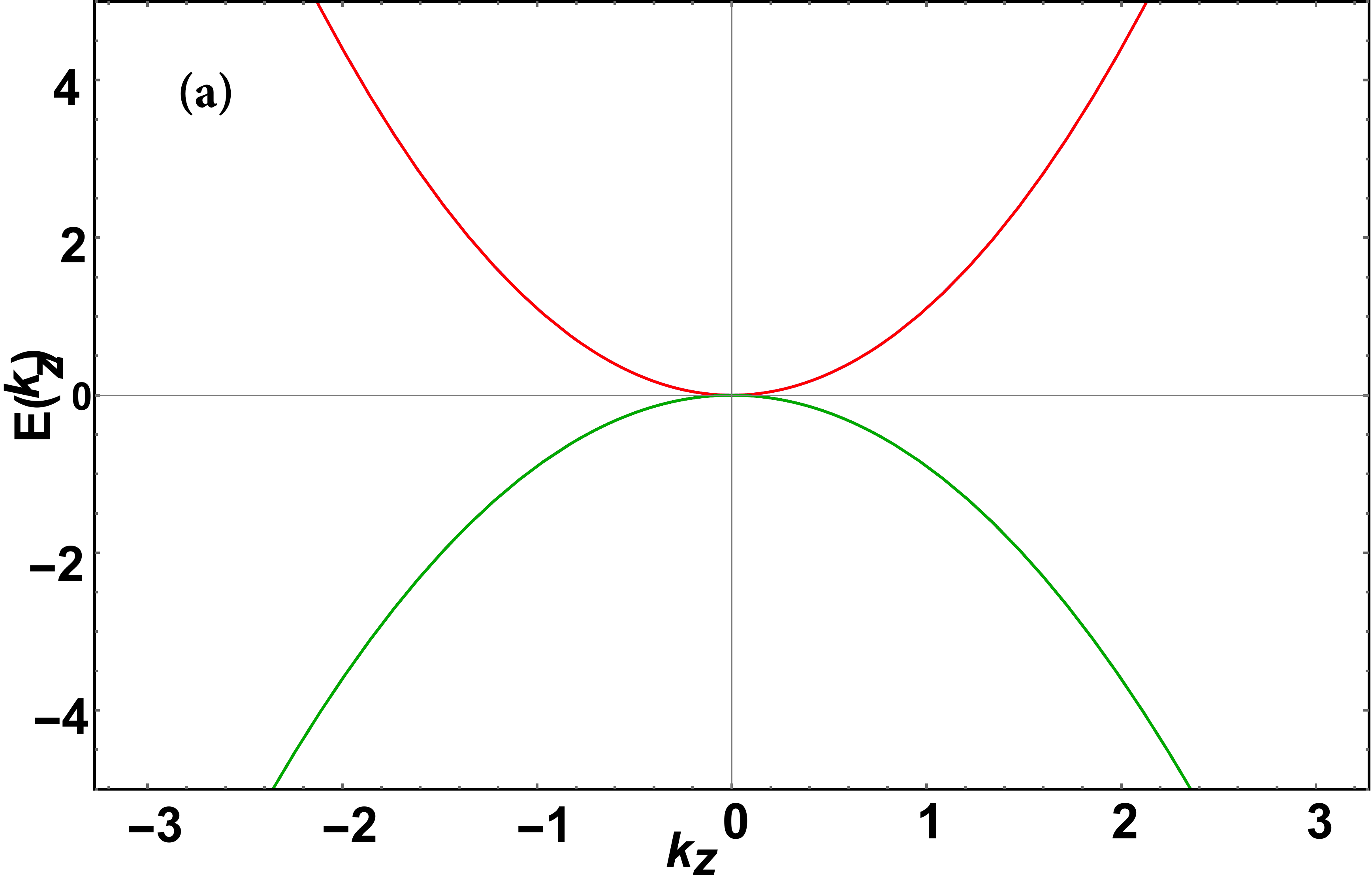}\includegraphics[width=0.25\textwidth]{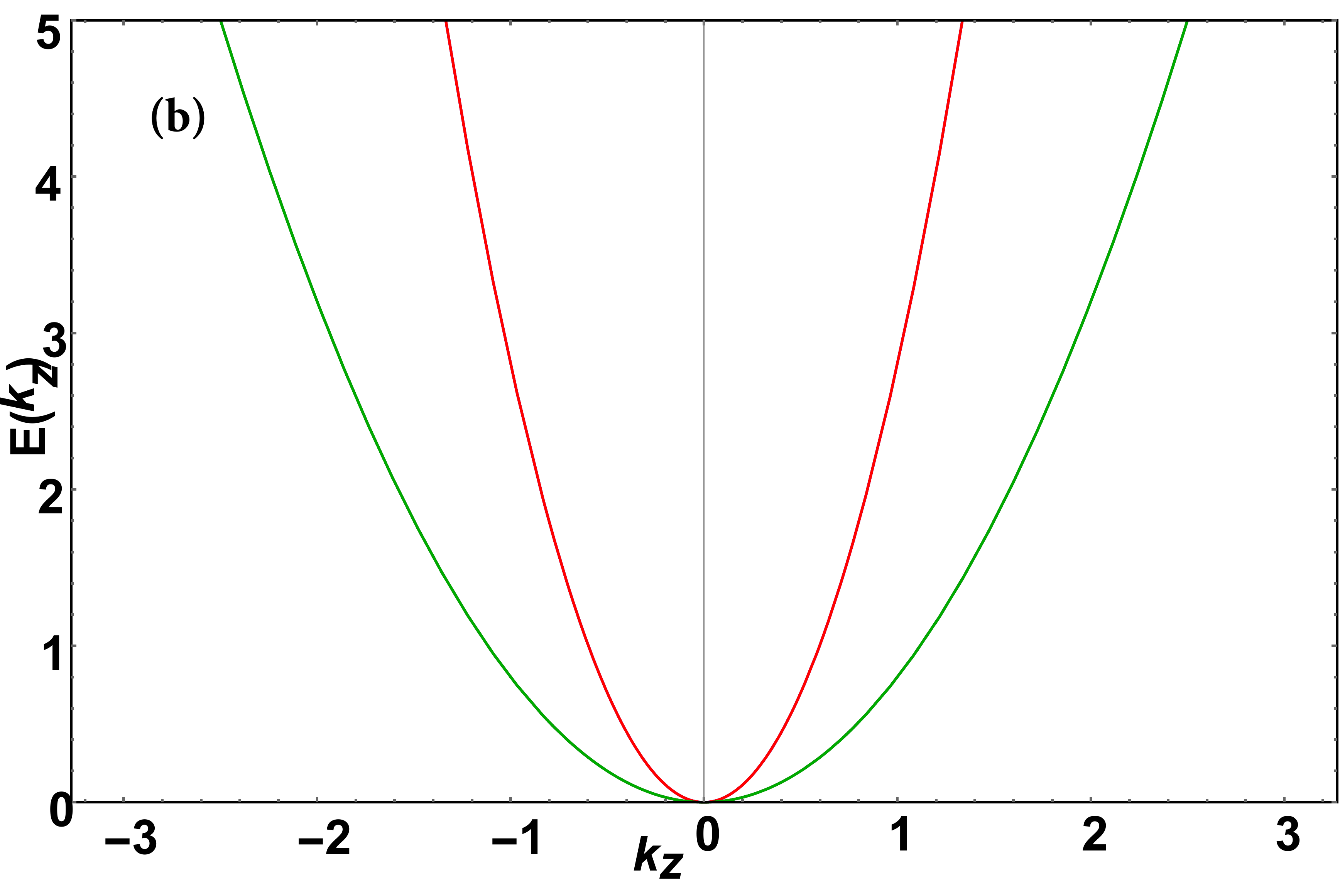}
\caption{Energy dispersion of Eq. (1) for (a) $\lambda_{1}=0.1$ and $\lambda_{2}=0.5$
with $J=3/2$ (red) bending down $J=1/2$ (blue) bending up, and (b)
$\lambda_{1}=1.8$ and $\lambda_{2}=0.5$ where both bands bend up.
Note that the bands are doubly degenerate.}
\end{figure}

Here, we study another aspect of this model. By employing machinery
from Floquet theory, we investigate light-matter interactions in this
model in both band-bending scenarios. We consider periodic driving
induced by laser light with a general vector potential $\vex{\mathcal{A}}(t)=(A_{x}\cos(\omega t),A_{y}\eta\sin(\omega t),0)$
, where $\eta=\pm1$ corresponds to right-handed and left-handed polarizations
of the light, respectively, and $A_{i}\propto E_{i}/\omega$, where
$E_{i}$ is its electric field. The time-dependent Hamiltonian can
be written as $H(\vex{k},t)=\sum_{n}H_{n}(\vex{k})e^{in\omega t}$,
where $H_{\pm n}(\vex{k})=\frac{1}{T}\int_{0}^{T}H(\vex{k},t)e^{\pm in\omega t}$.\\
 The effective time-independent Hamiltonian in the high frequency
limit, as dictated by Floquet theory, is \cite{27,28}, 
\begin{align}
H_{eff}(\vex{k})=H_{0}+\sum_{n\geq1}\frac{[H_{+n},H_{-n}]}{n\omega}+O\left(\frac{1}{\omega^{2}}\right).
\end{align}
where, 
\begin{align}
H_{1}= & \,(\lambda_{1}+\frac{5}{2}\lambda_{2})\vex{k}.\vex{A}-2\lambda_{2}\{\vex{J}.\vex{k},\vex{J}.\vex{A}\}\\
H_{2}= & \,\frac{1}{4}((\lambda_{1}+\frac{5}{2}\lambda_{2})\vex{A}^{2}-2\lambda_{2}(\vex{J}.\vex{A})^{2})\\
H_{-n}= & H_{n}^{\dagger},
\end{align}
and $\vex{A}=(A_{x},i\eta A_{y},0)$. The Floquet perturbation series
is controlled by parameter $\gamma=\lambda e^{2}E^{2}/\hbar\omega^{3}$,
where $\lambda$ is either $\lambda_{1}$ or $\lambda_{2}$ which
are of the same order of magnitude and have units of inverse mass,
$E$ is the magnitude of the electric field of the incident light
and $c$ is the speed of light in the medium. Clearly, $\gamma\ll1$
at high enough frequencies, thus controlling the Floquet expansion.
We discuss the estimation of this parameter in the real experiments
in the concluding section. In the meantime, we work in the units $e=\hbar=1$.
We first analyze the limit of circular polarization, which is the
only case can be fully studied analytically. Then, we analyze the
general case of elliptical polarization.\\
 \textit{Circularly polarized light}.- Since $H$ is quadratic in
$\vex{k}$, $H_{n}=0$ for $n>2$ in Eq. (2). The terms coming from
$n=\pm2$ are momentum-independent, and it is their competition with
the $\mathbf{k}$-dependent terms arising from $n=\pm1$ that proves
to be essential in realizing the various WSMs. In other words, the
leading order correction in $\vex{A}$ is insufficient, and it is
necessary to go to a higher order. For circularly polarized light,
rotational symmetry ensures that Weyl points appear only on the $k_{z}$
axis, which makes extracting the salient features of the model analytically
possible. For $k_{x}=k_{y}=0$, the effective Hamiltonian reads 
\begin{align}
H_{eff}(k_{z})=H_{0}(k_{z})+ & \,\frac{2i\eta A^{2}\lambda_{2}^{2}}{\omega}\bigg(-k_{z}^{2}[\{J_{x},J_{z}\},\{J_{y},J_{z}\}]\\
+ & \,\frac{A^{2}}{8}[\ J_{y}^{2}-J_{x}^{2},\{J_{x},J_{y}\}]\bigg),
\end{align}
with dispersions of $E_{1,\pm}=(\lambda_{1}+2\lambda_{2})k_{z}^{2}\pm(3A^{2}\lambda_{2}^{2}\eta(A^{2}-8k_{z}^{2}))/2\omega-\mu$
and $E_{2,\pm}=(\lambda_{1}-2\lambda_{2})k_{z}^{2}\pm(3A^{2}\lambda_{2}^{2}\eta(A^{2}+8k_{z}^{2}))/2\omega-\mu$.
Note that introduction of circularly polarized light has broken time-reversal
symmetry and lifted the double degeneracy of the bands. Inversion
symmetry survives, though, because only even powers of the light amplitude
enter $H_{eff}$. The four non-degenerate bands intersect in various
pairs, giving rise to Weyl nodes at $\vec{K}_{1}=(0,0,\pm A/2\sqrt{2})$
and $\vec{K}_{2}=(0,0,\mp A^{2}\sqrt{3\lambda_{2}/\omega}/2)$. We
can compute the monopole charge of each node by writing an effective
low energy Hamiltonian around in the form $H_{k}\propto\vex{n}(\mathbf{k})\cdot\vex\sigma$
and using, 
\begin{align}
W_{n}=\int_{S}d^{2}\vex{k}\epsilon^{ijk}\vex{n}.(\partial_{j}\vex{n}\times\partial_{k}\vex{n})
\end{align}
where $\vex{n}$ is a unit vector and the integration is over a surface
$S$ surrounding the node. We obtain $W_{n}=\pm1$ and $W_{n}=\pm2$
for $K_{1}$ and $K_{2}$ respectively. This is a remarkable result,
that single and double-Weyl nodes coexist at different energies, thus
allowing us to access both dynamically by tuning the chemical potential.
As is clear, the positions of single Weyl points are only a function
of the light parameters while the locations of the double-Weyl points
also depend on the band structure parameter, $\lambda_{2}$. Moreover,
for circularly polarized light, there is a special point in parameter
space, namely, $A_{m}=\pm\sqrt{\omega/6\lambda_{2}}$ where the two
types of nodes merge and form a triply degenerate point (TDP).\\
 
\begin{figure}
\includegraphics[width=0.5\textwidth]{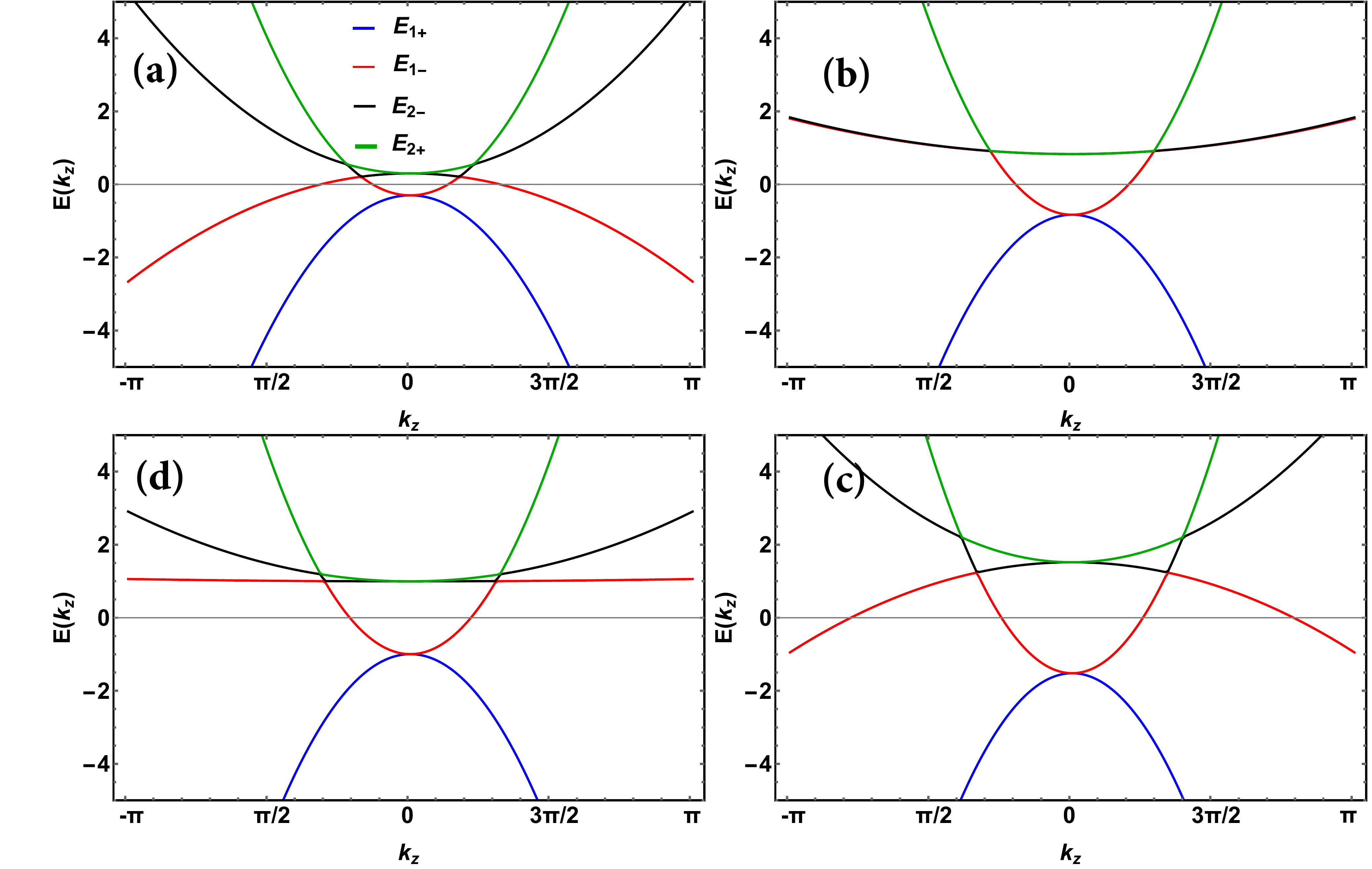}\\
 \includegraphics[width=0.5\textwidth]{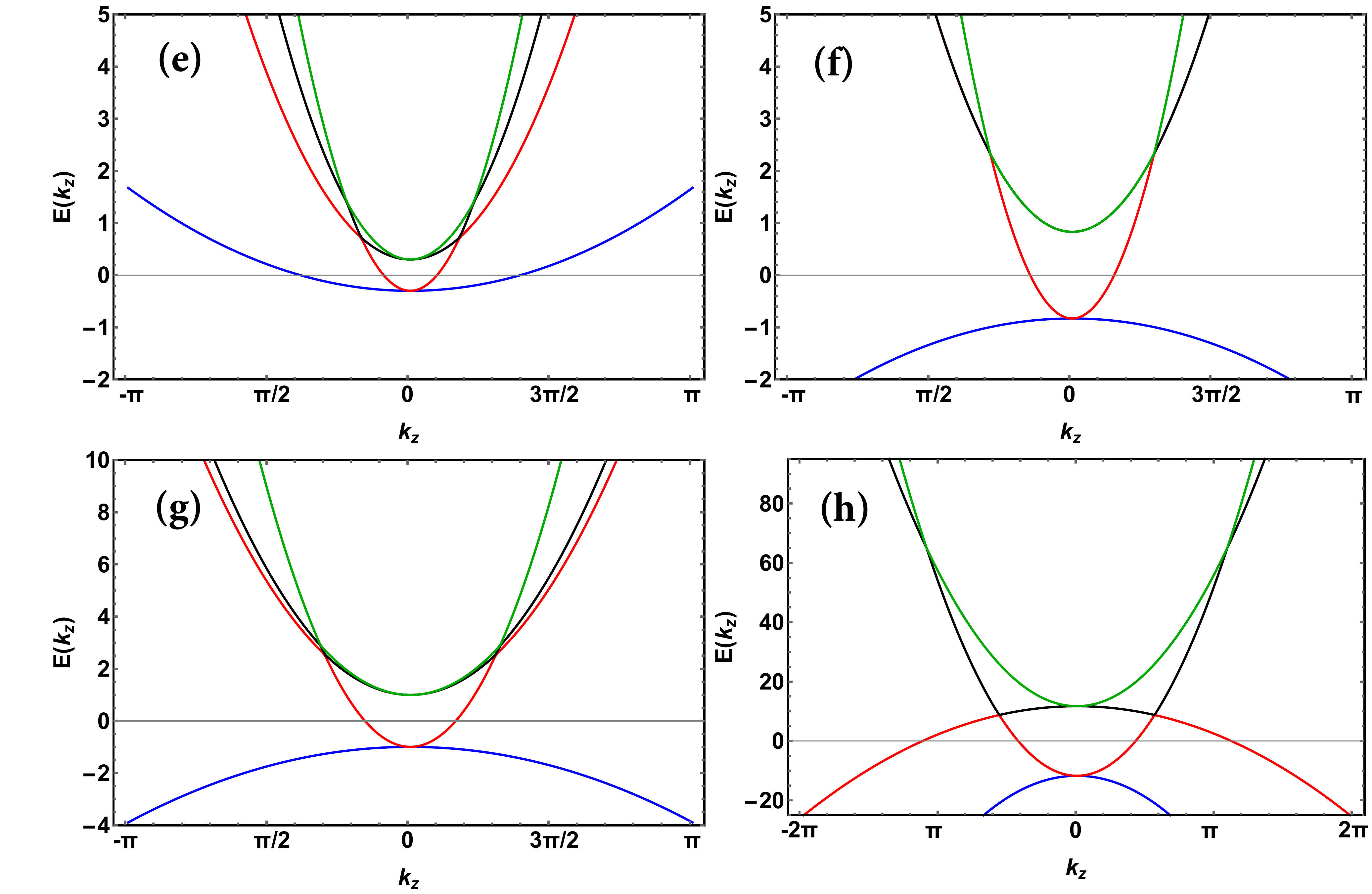} \caption{Evolution of the Weyl nodes with light amplitude $A$. (a)-(d) show
$A=2,2.58,2.7$ and $3$, respectively for bands bending oppositely.
For (e)-(h), we used $A=2,2.58,2.7$ and $5$, respectively, to show
the type-II to type-I phase transition for high enough intensity with
both bands bending in same direction. $\lambda_{1}=0.1,\lambda_{2}=0.5$
($\lambda_{1}=1.8,\lambda_{2}=0.5$) are used for bands bending oppositely
(similarly). $\omega=20$, $\mu=0$ and $\eta=1$ is used for all
of the plots.}
\end{figure}

$\,\,\,\,$Fig. 3(a-d) shows the evolution of the band structure with
the light intensity, for a representative set of parameters with $\eta\lambda_{2}>0$
in a scenario with bands bending oppositely. This corresponds to evolution
along the $A_{x}=A_{y}$ line in Fig. 1. Two pairs of nodes appear
(Fig. 3a) as soon as light is turned on. The nodes higher (lower)
in energy are type-II (type-I), have monopole charge $\pm1$ ($\pm2$)
and occur at $K_{1}$ ($K_{2}$). On increasing $A$, the lower nodes
flatten along $k_{z}$ (not shown) and transition into type-II nodes,
before merging with the upper nodes at the TDP at $A=A_{m}$ (Fig.
3b). On further increasing $A$, the bands cross, and the charge $\pm2$
nodes end up being higher in energy than the charge $\pm1$ nodes
(Fig. 3c). The latter then transitions back from type-II to type-I
(Fig. 3d). In summary, the upper nodes are always type-II, while the
lower nodes evolve from type-I to type-II and back to type-I. Naturally,
the bands hosting the lower Weyl nodes flatten twice during this evolution,
once at each transition between type-I and type-II characters. The
charges are $\pm1$ ($\pm2$) for the upper (lower) nodes for low
intensity, and the correspondence gets reversed as $A$ is tuned across
the TDP.

Fig. 3(e-h) show the evolution when the bare bands bend the same way.
It shows the same trend as the case where the bare bands bend oppositely,
except that all the nodes are type-II. Moreover, there is a type-II
to type-I transition at high intensity, as shown in Figs. 3(g) and
3(h).

\textit{Elliptically polarized light}.-Now, we turn to the more general
case of elliptically polarized light, i.e, $A_{x}\neq A_{y}$. The
phase diagram is much richer when the incident light is anisotropic
in the field's amplitudes. In the following, we analyze various driven
phases in the two band-bending possibilities shown in Fig. 2.

\begin{figure}
\includegraphics[width=0.5\textwidth]{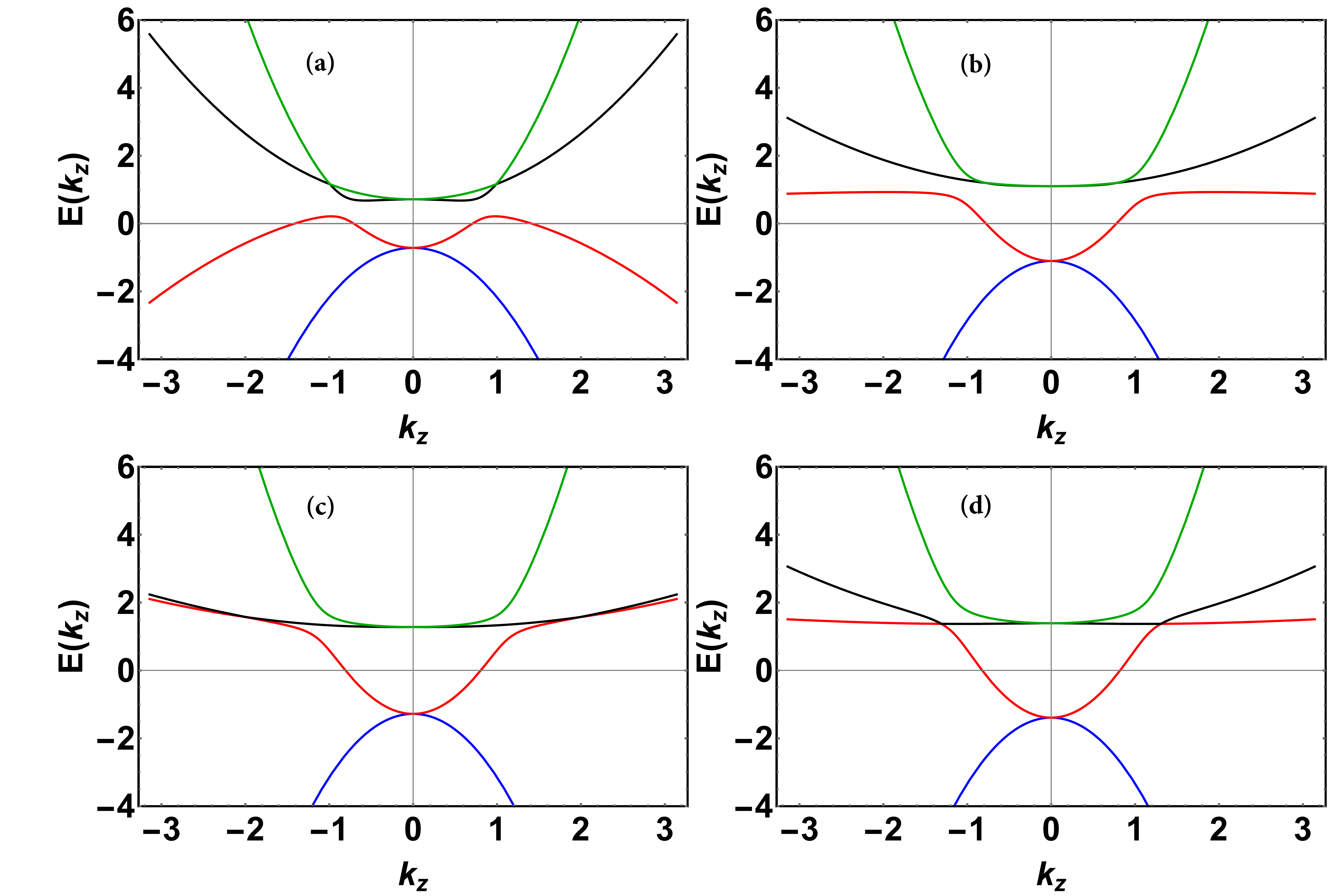}\\
 \includegraphics[width=0.27\textwidth]{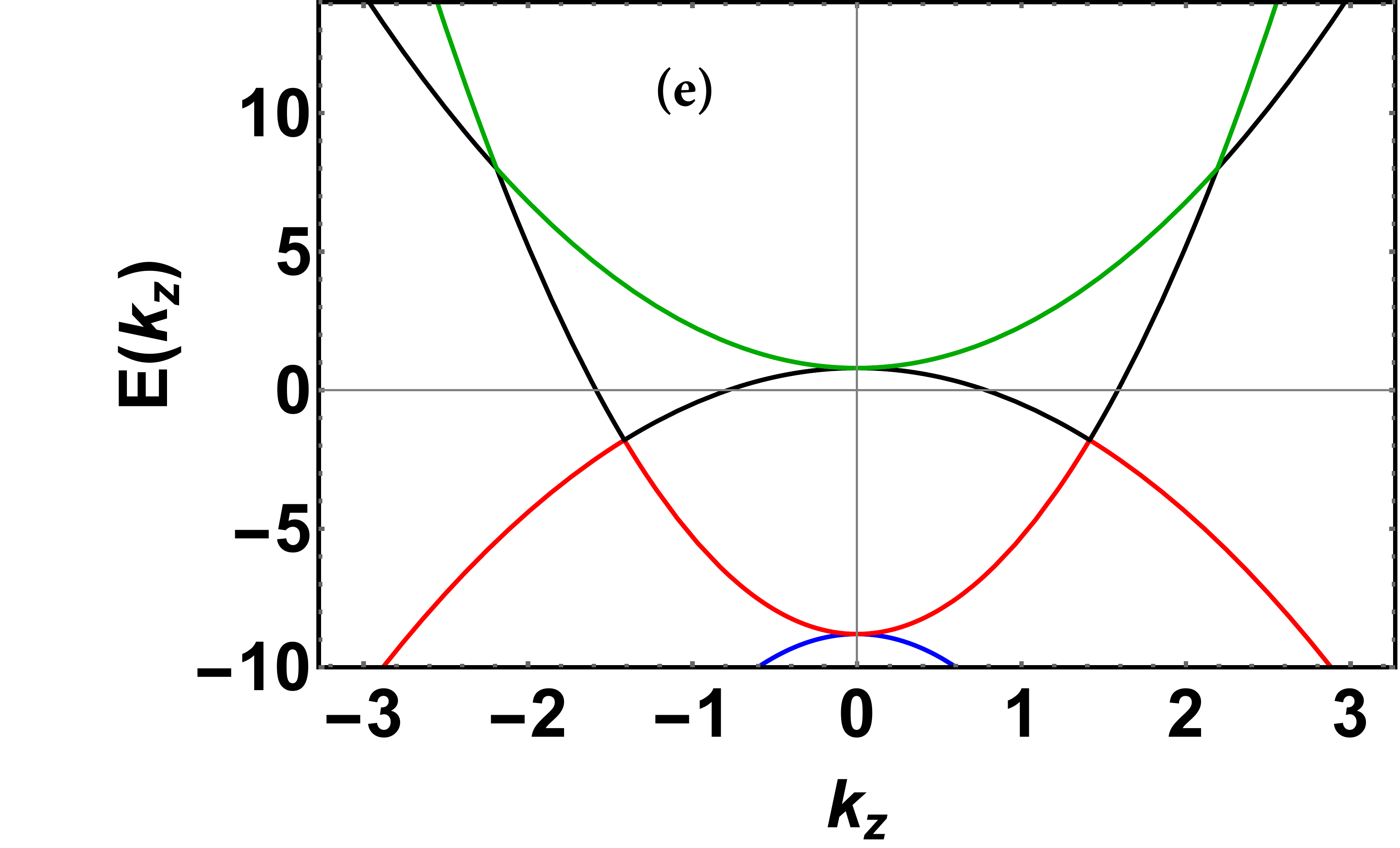} \caption{(a)-(e) show the $k_{x}=k_{y}=0$ cut for $A_{y}=4$ and $A_{x}=1,1.48,1.68,1.8,4$
respectively. $\lambda_{1}=0.1,\lambda_{2}=0.5$, $\omega=20$, $\eta=1$
and $\mu=0$ ($\mu=4$) for (a)-(d)((e))}
\end{figure}

\begin{figure}
\includegraphics[width=0.4\textwidth]{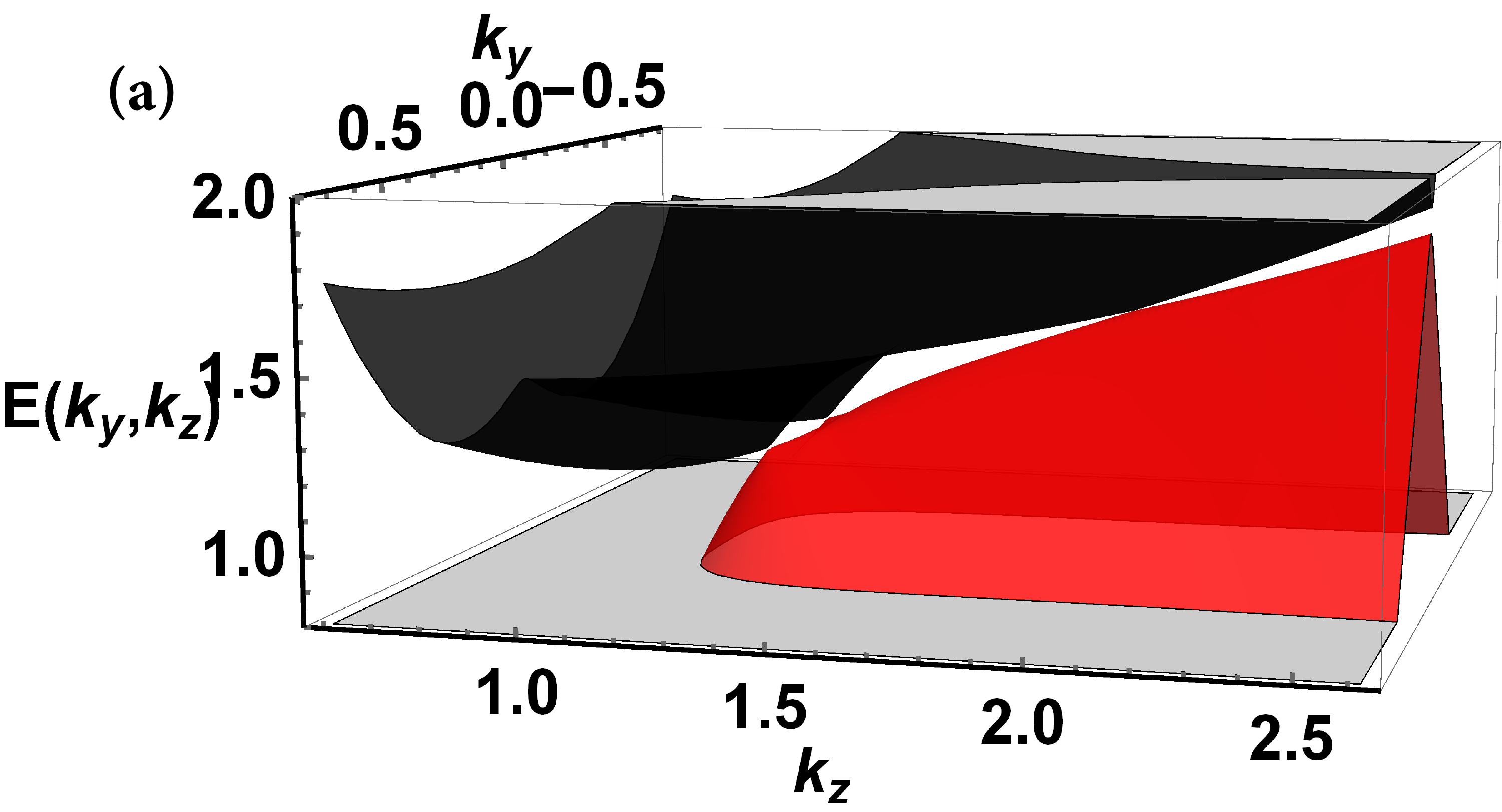}\\
 \includegraphics[width=0.4\textwidth]{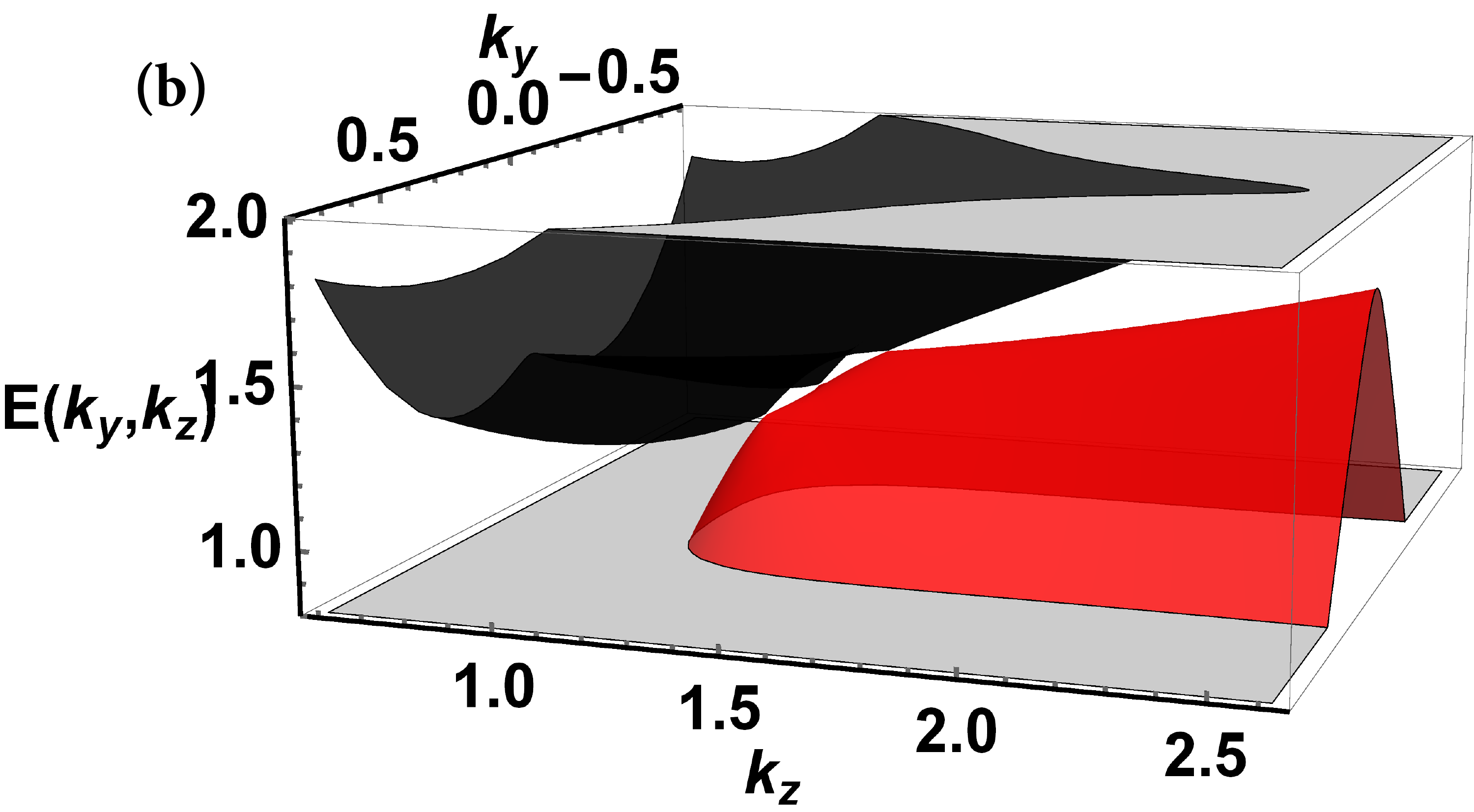}\\
 \includegraphics[width=0.4\textwidth]{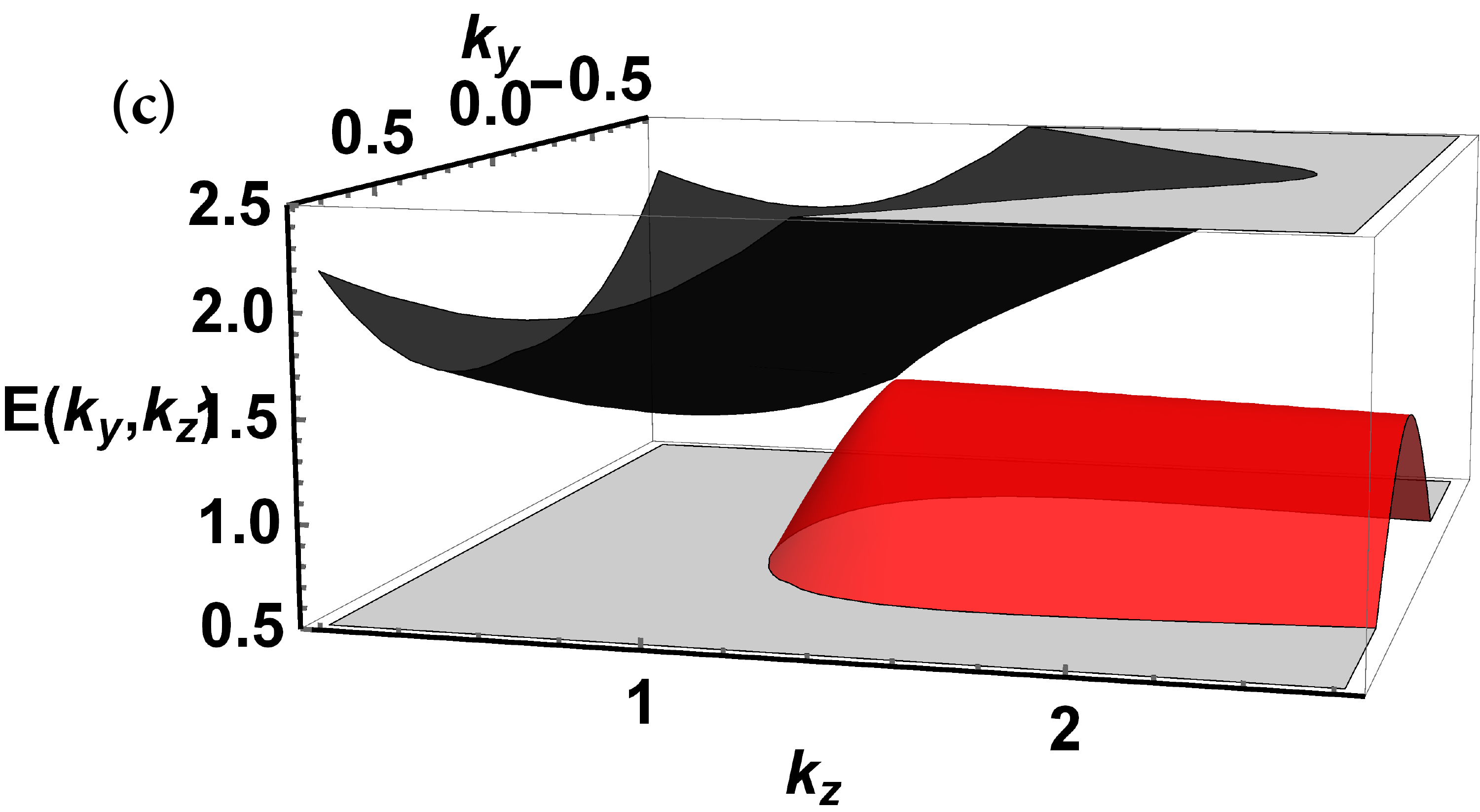} \caption{Appearance of type-II node from infinity (a) while there are two nodes
on the $k_{y}-k_{z}$ plane. The node from infinity moves towards
origin (b) and merges with two other nodes on the $k_{z}$ axis (c).
(a)-(c) we used fixed $A_{y}=4$ and $A_{x}=1.68,1.72,1.85$, respectively.
To increase the resolution only evolution of the nodes between two
relevant bands is depicted. Also, only one side of plot is shown.}
\end{figure}

\begin{figure}
\includegraphics[width=0.4\textwidth]{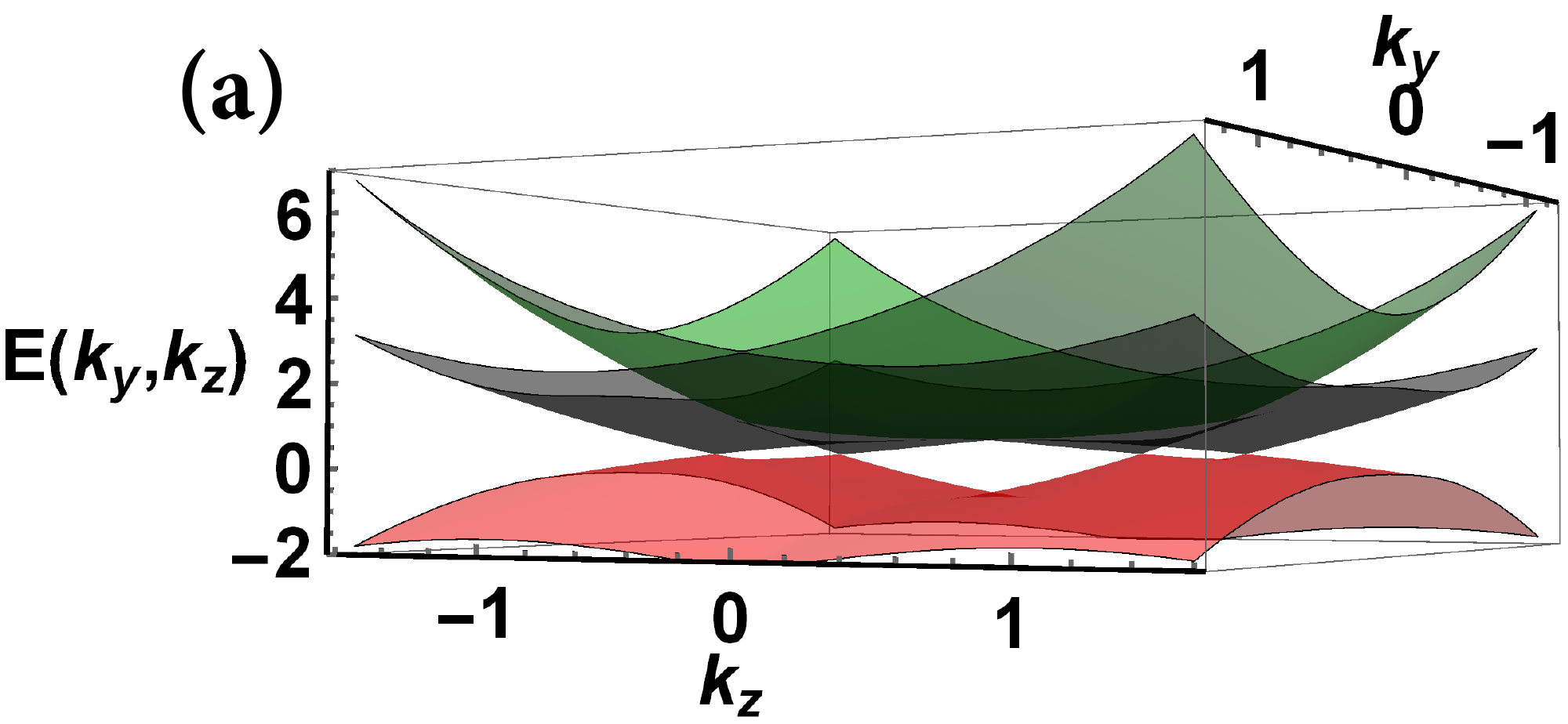}\\
 \includegraphics[width=0.35\textwidth]{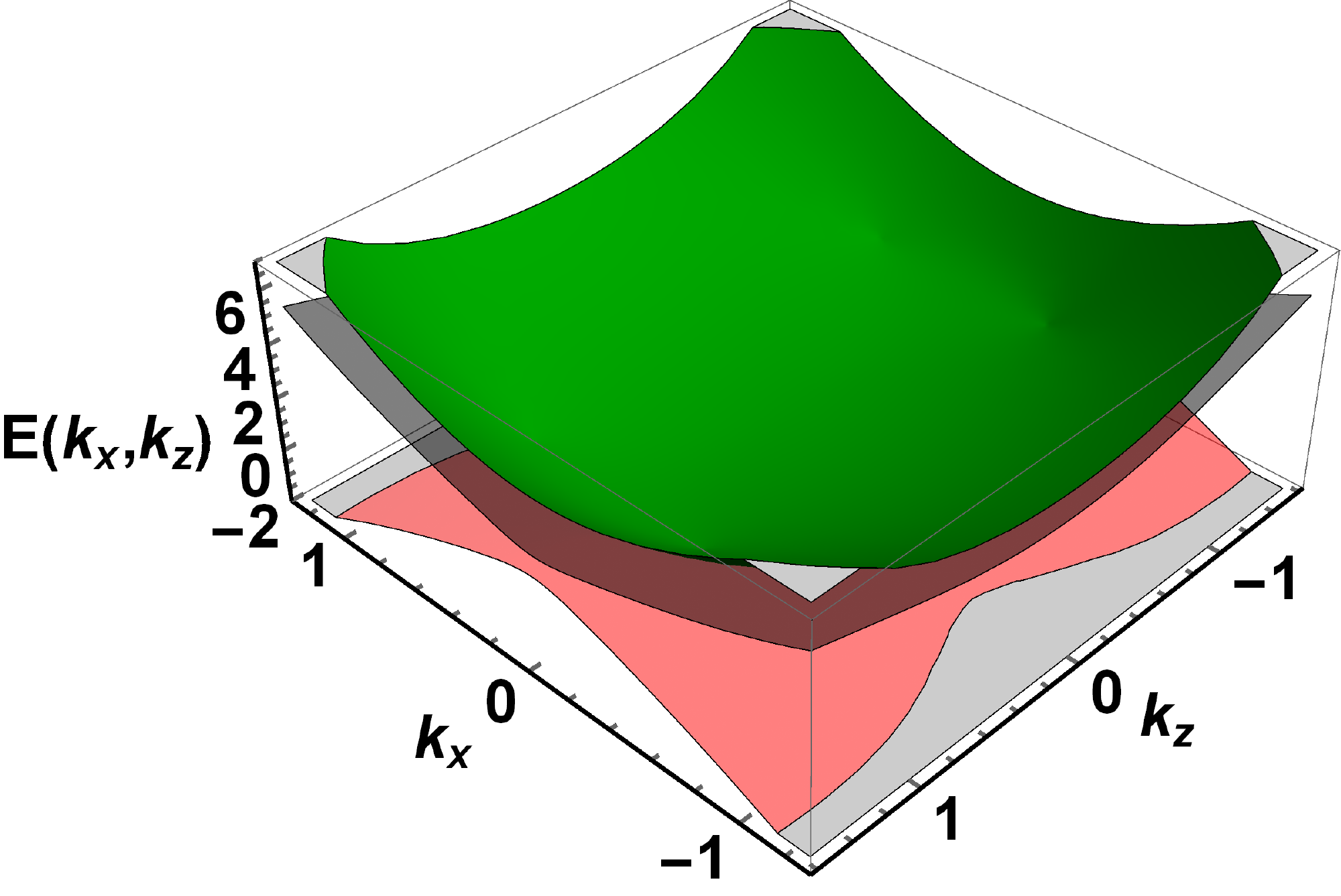}\\
 \includegraphics[width=0.35\textwidth]{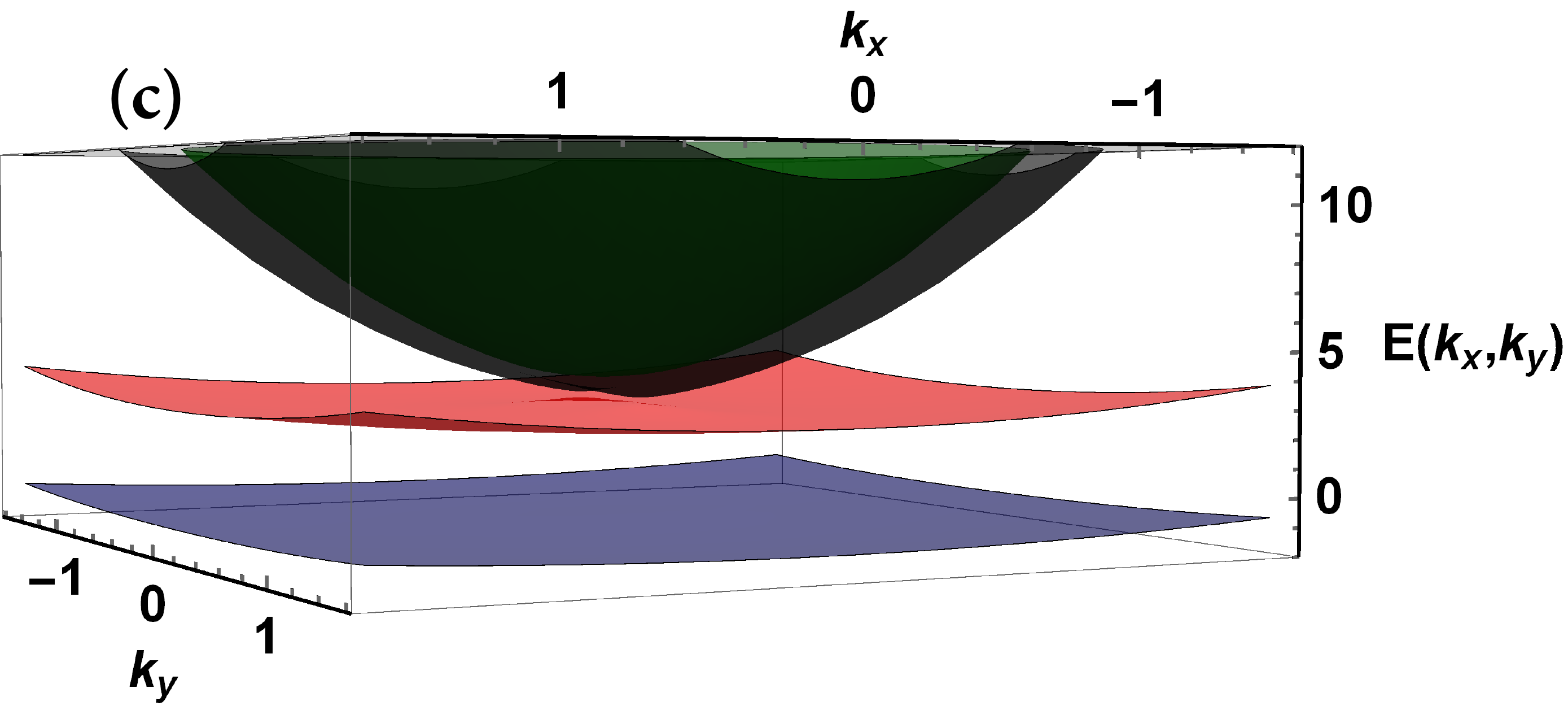} \caption{Representative 3D plots: (a) shows a representative plot in phase
(I) illustrating four nodes in the $k_{y}-k_{z}$ plane with $A_{y}=4$
and $A_{x}=1$, (b) shows a representative plot in phase (III) illustrating
four nodes in $k_{x}-k_{z}$ plane with $A_{y}=4$ and $A_{x}=2$.
Red rings denote nodes in the upper bands. (c) shows a representative
plot of the flat-bands in $k_{x}-k_{y}$ in the scenario with both
bands bending in same direction (Fig. 2b) with $A_{y}=4$ and $A_{x}=1.6$.
We used $\omega=20$, $\eta=1$ and $\mu=0$ for (a)-(c).}
\end{figure}

Let us first study Eq. (4) when the bands are bent oppositely. The
phase evolution for this case is depicted in Fig. 4. In describing
the evolution, we keep $A_{y}$ fixed at a high or a low value, and
tune $A_{x}$ from $0$ to $A_{y}$.

Let us first look at high $A_{y}$. For large anisotropy with $A_{y}\gg A_{x}$,
there are 4 type-I nodes of unit monopole charge in the $k_{y}-k_{z}$
plane and 2 type-II nodes of charge $\pm2$ at higher energies on
the $k_{z}$-axis (Fig. 4a and 6a). On increasing $A_{x}$, the two
higher nodes split into four type-II nodes of unit charge in the $k_{x}-k_{z}$
plane (Fig. 4b and 6b). On further increasing $A_{x}$ a pair of type-II
nodes of unit charge come in on the $k_{z}$ axis from $k_{z}=\pm\infty$
while there are still 4 nodes in $k_{y}-k_{z}$ plane (Fig. 4c and
5a). The new node at $k_{z}>0$ presumably has a monopole charge opposite
to that of the two nodes in the $k_{y}-k_{z}$ plane at $k_{z}>0$;
an analogous condition holds for $k_{z}<0$ with all monopole charges
reversed. Finally, the nodes in each triplet merge to yield two type-II
nodes of charge $\pm1$ (Fig. 5) on the $k_{z}$ axis. These nodes
change character from type-II to type-I, accompanied by the flattening
of one of the bands participating in the nodes (Fig. 4d), and survive
in this form up to the circularly-polarized limit (Fig. 4e). In the
meantime, the four higher nodes remain type-II with unit charge, but
merge into two type-II, charge $\pm2$ nodes in the limit of $A_{x}=A_{y}$.
Therefore we end up of 2 higher (type-II, $\pm2$) and lower (type-I,
$\pm1$) nodes as explained in the previous section (Fig. 4e). It
should be noted that the TDPs are absent for elliptical polarization.\\
 However, for lower $A_{y}$, situation is different. In this case
the situations of Fig. 4(b-d) do not happen. In another words, for
lower $A_{y}$, the upper nodes do not split, while the merging of
lower four points happens near the isotropic limit. Moreover, no flat-line
occurs, and the upper nodes (type-II) have charge $\pm1$ and lowers
(type-I) have $\pm2$ charges.

When both bare bands bend in same direction, the phase diagram undergoes
almost the same evolution as the case with bands bent oppositely.
In particular, it starts with four lower and two higher nodes, which,
after a series of merging and splitting, yields a phase with two lower
and four higher nodes. Finally, in the circular polarization limit,
the higher nodes merge, leaving only two lower and two higher nodes.
However, there are couple of differences. Firstly, as we mentioned
in the previous section, nodes are type-II for most amplitude ranges.
The second, instead of a flat line along $k_{z}$, there is a \char`\"{}flat-band\char`\"{}
in the $k_{x}-k_{y}$ plane for the lower nodes (Fig. 6c). The flat-line
along $k_{z}$ does happen, but only for very large $A_{y}$. This
is consistent with the type-II to type-I transition that was found
to occur at high intensities in the isotropic limit for bare bands
bending in the same direction.\\
 \textit{Discussion, experimental considerations and concluding remarks}.-
In this work we have studied the Floquet theory of the three-dimensional
Luttinger semimetal with quadratic band touching points. We have found
that depending on the orientation of bands and light parameters, both
type I and II Weyl nodes with single and double monopole charges at
different energies can be generated. In particular we arrive at the
following main results: 
\begin{itemize}
\item When the incident light is circularly polarized, we have solved the
problem analytically and have obtained two nodes with charge $\pm1$
and two other nodes with $\pm2$ at different energies. For both band
bending scenarios, the higher nodes are always type-II while the lower
ones can be type-I or II depending on light parameters. We found that
at a certain light intensity, both pairs of nodes merge to form two
TDPs. This is a special point which exists only for circularly polarized
light and is a function of both light and band structure parameters. 
\item For the elliptically polarized light, we have only solved system numerically.
For both bands bending scenarios, for large anisotropy, $A_{x}\ll A_{y}$,
there are two higher nodes on the $k_{z}$ axis and four lower nodes
in $k_{y}-k_{z}$ plane. Then, when the $A_{y}$ is held fixed at
a small value and $A_{x}$ is increased, the four lower nodes merge
around $A_{x}\sim A_{y}$. However, for high enough $A_{y}$, on increasing
$A_{x}$, the lower Weyl nodes merge and then tilt back and turn into
a flat-line (flat-band) for bare bands bending oppositely (similarly)
and make two nodes. On the other hand the higher nodes deform to nearly
flat lines and then split to four nodes in $k_{x}-k_{z}$ plane which
finally merge at isotropic limit. 
\end{itemize}
Therefore, we conclude that the Luttinger semimetal with parabolic
dispersion provides a master platform for realizing various types
of WSMs from type I to type II with four nodes or two nodes, as well
as single and double monopoles. Remarkably, we found that single and
double-Weyl can coexist at different energies, so either ones can
be accessed through controlled doping of the system and tuning of
laser light. To the best of our knowledge this is the only system
that reported so far with this level of tunability for broad range
of possible Weyl phases. In addition, irradiated Luttinger semimetals
is the only example so far discovered with Weyl node with different
monopole charges coexist, making it feasible for the possible applications
of both single and double WSMs. To the best of our knowledge our work
is the only example that can generate such a broad range of WSMs from
a system with no Weyl nodes. There have been some recent studies \cite{rec2,rec3},
where photo-induced multi-Weyl phases were generated from crossing-line
systems.

The Luttinger Hamiltonian describes a wide range of materials from
semiconductors to pyrochlore iridates and half-Heuslers which are
accessible experimentally, unlike the other semimetals such as Dirac,
loop-node, or linked semimetals, where experimental examples are rare
or non-existent. Therefore, this work might facilitate the experimental
realizations of photoinduced WSMs. Using $\lambda_{2}=4.2/m_{0}$
for HgTe, where $m_{0}$ is the bare electron mass, $\hbar\omega=120meV$
and an electric field of $E_{0}=2.5\times10^{7}V/m$ \textendash{}
typical values for pump-probe experiments \cite{pump} \textendash{}
we estimate the perturbation parameter $\gamma=\lambda e^{2}E^{2}/\hbar\omega^{3}\sim10^{-10}$,
so the Floquet expansion is certainly well-controlled. The only word
of caution is that, as with all three-dimensional Floquet systems,
our proposal only works for films thin enough for the electric field
to penetrate the system substantially.

%%%%%%%%%%%%%%%%%%%%%%%%%%%%%%%%%%%%%%%%%%%%%%%%%%%%%%%%%%%%%%%%%%%%%%%%%%%

\section*{ACKNOWLEDGEMENTS}

We would like to thank Bitan Roy, Matthew. S Foster and P.N. Ong for fruitful
discussions. S.A.A.G. and C.S.T. was supported by Texas Center for
Superconductivity and the Welch Foundation Grants No. E-1146. P. H.
was supported by the Division of Research, College of Natural Sciences
and Mathematics and the Department of Physics and the University of
Houston.

\end{document}